\documentclass[aps,prl,reprint,superscriptaddress,nofootinbib,longbibliography]{revtex4-2}

\usepackage[colorlinks=true,linkcolor=blue,citecolor=blue,urlcolor=blue,filecolor=blue]{hyperref}
\usepackage{graphicx}
\usepackage{amsmath}
\usepackage{amssymb}
\usepackage{amsfonts}

\newcommand{\rnhead}[1]{\par\medskip\noindent\textit{#1}.---\hspace{0.4em}\ignorespaces}

\begin{document}

\title{Heat-current-induced nonreciprocal phonon damping}

\author{Tingwei Deng}
\thanks{These authors contributed equally to this work.}
\affiliation{National Laboratory of Solid State Microstructures and School of Physics, Nanjing University, Nanjing 210093, China}
\affiliation{Collaborative Innovation Center of Advanced Microstructures, Nanjing University, Nanjing 210093, China}
\author{Xianyong Ding}
\thanks{These authors contributed equally to this work.}
\affiliation{College of Physics and Optoelectronic Engineering, Chongqing Normal University, Chongqing 401331, China}
\author{Rui Wang}
\thanks{Corresponding author: rcwang@cqu.edu.cn}
\affiliation{Institute for Structure and Function and Department of Physics, Chongqing University, Chongqing 400044, People's Republic of China}
\affiliation{Center for Quantum Materials and Devices, Chongqing University, Chongqing 400044, People's Republic of China}
\author{Xiangang Wan}
\thanks{Corresponding author: xgwan@nju.edu.cn}
\affiliation{National Laboratory of Solid State Microstructures and School of Physics, Nanjing University, Nanjing 210093, China}
\affiliation{Collaborative Innovation Center of Advanced Microstructures, Nanjing University, Nanjing 210093, China}


\begin{abstract}
Time-reversal symmetry enforces equal equilibrium linewidths for time-reversed
phonon modes at opposite wave vectors.
Here we show that a steady heat current can produce nonreciprocal
anharmonic damping even in centrosymmetric nonmagnetic crystals.
The effect arises from nonequilibrium occupations of scattering partners,
which change the statistical weights of time-reversed three-phonon
channels while their squared matrix elements remain equal.
Within the single-mode relaxation-time approximation, we derive a linewidth
correction that is linear in the temperature gradient and odd under
wave-vector reversal.
The response depends on scattering-partner velocities and equilibrium
lifetimes and retains signed channel weights absent from the equilibrium linewidth.
First-principles calculations for centrosymmetric MgO and diamond yield
nonzero, mode-dependent linewidth corrections at selected finite wave vectors,
demonstrating the mechanism in both ionic and covalent crystals.
Reversing the heat current at fixed local temperature exchanges the
broadening and narrowing of paired modes.
This current-reversible linewidth contrast provides a spectral signature
of occupation feedback, with potential implications for nonlinear phonon transport.
\end{abstract}

\maketitle

\rnhead{Introduction}
Understanding heat transport is essential for controlling energy flow in
solids and managing heat in electronic devices
\cite{WOS:000348494300006,WOS:000372424100001,Chen2022Interfacial}.
Anharmonic scattering connects phonon heat transport with the decay of
lattice vibrations \cite{BornHuang,Ziman,Peierls,WOS:A19621485C00036,Cowley}.
Phonon lifetimes and spectral linewidths reveal the microscopic
interactions behind this connection~\cite{WOS:A1995RQ99000035,WOS:A19668114800046,WOS:A1984SD22400056,Liu2024Lifetime}.
Harmonic and cubic force constants from first-principles calculations
enable quantitative investigations of lattice thermal transport~\cite{WOS:000255457300022,WOS:000294227700004,WOS:000270383300059,Luo2025Tensor}.
Together with experiments, these calculations link anharmonic scattering to
thermal properties across a broad range of materials
\cite{WOS:000384537600001,WOS:000458767000011,WOS:000251450600044,WOS:A1990DP62800002,Zhou2024UO2}.
Because heat-carrying phonons also serve as scattering partners, their
nonequilibrium occupations can modify phonon damping. Quantifying this feedback requires connecting the steady heat-carrying
distribution to the occupation dependence of phonon decay. 
Iterative Boltzmann methods and more general transport formulations
describe the heat-carrying distribution
\cite{WOS:A1995RK51900001,WOS:000885293700001}.
Occupation-dependent self-energies link phonon populations to
coherent-phonon damping and nonequilibrium anharmonic dynamics after optical excitation
\cite{Pan2025,Mocatti2026}.
Directional phonon populations can also modify electronic scattering and
nonlinear photocurrents \cite{Xu2022PhononDressing}.

At equilibrium, time-reversal symmetry requires equal linewidths for
time-reversed modes at opposite wave vectors. Breaking time-reversal symmetry
by magnetic order can enable nonreciprocal acoustic absorption through
magnetoelastic coupling
\cite{Sasaki2017,Liao2024}.
Directional driving offers another route by creating a nonequilibrium
state that is not time-reversal invariant, without requiring magnetic order.
Optical driving can give oppositely propagating phonons different damping
rates \cite{Kim2017ChiralCooling,Otterstrom2018}.
Drifting electrons can amplify acoustic phonons through
electron--phonon interactions \cite{Andersen2019,BarajasAguilar2024}. These advances show how optical or electronic driving modifies phonon damping. It remains unclear, however, whether a steady phonon heat current can produce
unequal intrinsic linewidths for time-reversed modes through the
nonequilibrium occupations of their anharmonic scattering partners. 


Addressing this question requires distinguishing changes in the local
anharmonic linewidth from other spectral effects of heat flow.
The quantity of interest is the linewidth contrast between time-reversed
bulk phonon modes at fixed local temperature and heat current.
Early investigations focused on Stokes/anti-Stokes intensity asymmetries
and nonequilibrium Brillouin spectra
\cite{WOS:A1968C494800016,WOS:000073327500010,Beysens1980,Tremblay1981}.
Temperature gradients also induce Raman line-shape modifications via
spatial temperature variations \cite{WOS:A1984RZ08300046} and
occupation-dependent spectral skewness \cite{Rasolt1983}.
Neither intensity asymmetries nor line-shape distortions alone establish
an intrinsic linewidth contrast.

\begin{figure}[!t]
\includegraphics[width=\linewidth]{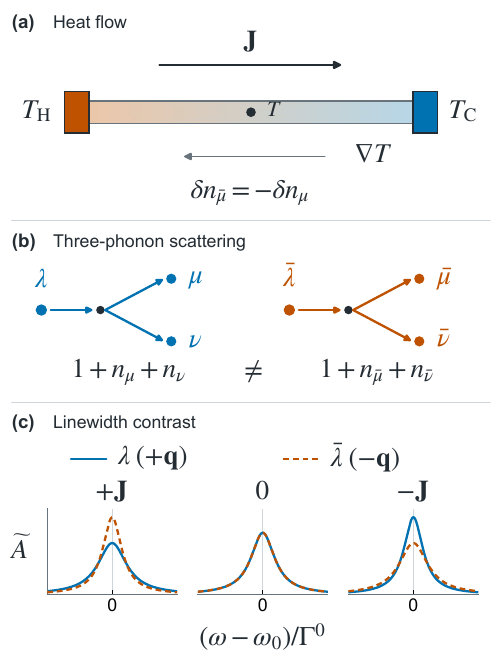}
\caption{Illustration of heat-current-induced nonreciprocal phonon damping.
(a) A temperature bias drives a steady phonon heat current in a
nonmagnetic crystal with time-reversal symmetry at equilibrium.
At local temperature $T$, time-reversed partners have opposite
first-order occupation corrections, $\delta n_{\bar\mu}=-\delta n_\mu$.
(b) Paired three-phonon decay channels have equal squared cubic matrix
elements but different partner occupations.
(c) Schematic Lorentzian profiles for $\lambda=(\mathbf q,j)$ and
$\bar\lambda=(-\mathbf q,j)$.
The paired linewidth contrast changes sign under current reversal,
$\Delta\Gamma_\lambda(-J)=-\Delta\Gamma_\lambda(J)$.}
\label{fig:mechanism}
\end{figure}

In this Letter, we show that heat-carrying phonons modify the damping
of a probe mode through their nonequilibrium occupations as scattering
partners. This occupation feedback can produce nonreciprocal anharmonic
damping even in centrosymmetric nonmagnetic crystals.
By comparing time-reversed modes at the same local temperature, we
separate the first-order, wave-vector-odd linewidth correction from
symmetric thermal broadening.
As shown in Fig.~\ref{fig:mechanism}(a), time-reversed scattering partners
acquire opposite first-order occupation corrections.
These corrections change the occupation factors of paired three-phonon
channels while their squared cubic matrix elements remain equal
[Fig.~\ref{fig:mechanism}(b)].
The time-reversed probe modes therefore acquire opposite first-order
linewidth corrections, which exchange upon current reversal at fixed
local temperature [Fig.~\ref{fig:mechanism}(c)].
To quantify this response, we evaluate the anharmonic self-energy with
first-order occupation corrections obtained within the single-mode
relaxation-time approximation (RTA).
This yields a relaxation-time correction that is linear in the
temperature gradient and feeds back into the phonon distribution at
second order in the gradient.
Equilibrium time-reversal symmetry fixes the wave-vector-odd form of
the first-order linewidth correction, but occupation feedback itself
does not require that symmetry.
First-principles calculations for MgO and diamond, and for the
phonon--phonon channel of TaAs, illustrate this response at selected
finite wave vectors.
This current-reversible linewidth contrast connects phonon heat
transport to directional damping through the same anharmonic interactions.

\begin{table*}[!t]
\caption{Phonon modes satisfying the first-order criterion at $T=50$ K and
$\nabla_x T=10^3$ K/m. The columns list the equilibrium lifetime $\tau^0$, first-order
heat-current correction $\tau_x^1=\delta\tau$ for an $x$-directed gradient,
coefficient $\gamma^x$, and relative
correction $\delta\tau/\tau^0$. Coordinates are in reciprocal lattice
units, and $\tau^0$ is evaluated in the three-phonon channel.
The TaAs values include only this phonon--phonon contribution.
The corrections are RTA estimates at the stated finite smearing widths.
The mode labels identify the same modes in the figures.}
\label{tab:tau}
\centering
\scriptsize
\begin{tabular}{llllccc}
\hline
mode/material & $\mathbf q$ & $\omega$ (THz) & $\tau^0$ (ps) & $\tau_x^1$ (ps)
& $\gamma^x$ (\AA/(K$\cdot$ps)) & $\delta\tau/\tau^0$ \\
\hline
MgO (M1) & $(0,0,-0.1875)$ & 7.03 & 298.6 & $-46.3$ & $-2598$ & $-15.5\%$ \\
MgO (M2) & $(0.375,0.375,0.75)$ & 14.81 & 5.04 & $-0.13$ & $-24668$ & $-2.5\%$ \\
diamond (C1) & $(0.25,0,0)$ & 20.77 & 75.6 & $-2.00$ & $-1751$ & $-2.65\%$ \\
diamond (C2) & $(0.4375,0,0)$ & 37.99 & 12.2 & $-0.067$ & $-2267$ & $-0.55\%$ \\
TaAs (T1) & $(0,0,-0.3125)$ & 2.55 & 1540.8 & $162.5$ & $342.2$ & $+10.5\%$ \\
TaAs (T2) & $(0,0,-0.375)$ & 8.16 & 15.8 & $0.20$ & $3974$ & $+1.3\%$ \\
\hline
\end{tabular}
\end{table*}

\rnhead{Mechanism and linewidth response}\label{sec:theory}
The mechanism follows from the occupation dependence of three-phonon
scattering.
As shown in Fig.~\ref{fig:mechanism}, we denote the probe mode by $\lambda=(\mathbf q,j)$, with time-reversed
partner $-\lambda\equiv\bar\lambda=(-\mathbf q,j)$.
The scattering partners $\lambda'$ and $\lambda''$ have occupations
$n_{\lambda'}$ and $n_{\lambda''}$.
A decay channel carries the factor $1+n_{\lambda'}+n_{\lambda''}$,
whereas a difference channel carries $n_{\lambda'}-n_{\lambda''}$.
Time-reversed channels have equal squared matrix elements, but a
heat-carrying distribution gives their scattering partners different
occupations.
We obtain the resulting linewidth response from the cubic anharmonic
Hamiltonian and Fermi's golden rule
\cite{BornHuang,WOS:A19621485C00036,Srivastava}.
The detailed derivation is given in Sec.~I of the Supplemental Material
(SM)~\cite{SupplementalMaterial}, and Sect.~III of the SM illustrates the mechanism with one pair of scattering partners.

Within the single-mode RTA
\cite{Peierls,Ziman,WOS:A1959WB76500017,Klemens},
we expand the phonon occupation as
$n_\lambda=n_\lambda^0+n_\lambda^1+n_\lambda^2$
through second order in the temperature gradient.
The first-order correction is
$n_\lambda^1=-\tau_\lambda^0
(\partial n_\lambda^0/\partial T)\,
\mathbf v_\lambda\cdot\nabla T$,
while $n_\lambda^2\propto(\nabla T)^2$.
Here $n_\lambda^0$ is the Bose--Einstein occupation at the local temperature.
The quantities $\tau_\lambda^0$ and $\mathbf v_\lambda$ denote the
equilibrium relaxation time and phonon group velocity, respectively.

We define the phonon linewidth $\Gamma_\lambda$ as the half-width at half maximum.
Its linear response at fixed local temperature is characterized by
$\gamma_\lambda^c=-[\partial\Gamma_\lambda/\partial(\nabla_c T)]_{\nabla T=0}$,
where $c$ labels a Cartesian direction.
Substituting the first-order occupation corrections of the scattering
partners into the three-phonon scattering rate gives
\begin{equation}
\resizebox{\columnwidth}{!}{$
\begin{aligned}
\gamma_\lambda^c
&=\frac{18\pi}{\hbar^2}\sum_{\lambda'\lambda''}
|\Phi_{-\lambda\lambda'\lambda''}|^2\!\times\Big\{
\Big[
\tau_{\lambda'}^0 v_{\lambda'}^c
\frac{\partial n_{\lambda'}^0}{\partial T}
+\tau_{\lambda''}^0 v_{\lambda''}^c
\frac{\partial n_{\lambda''}^0}{\partial T}
\Big]
\\&\times\delta(\omega-\omega_{\lambda'}-\omega_{\lambda''})
\\&+\Big[
\tau_{\lambda'}^0 v_{\lambda'}^c
\frac{\partial n_{\lambda'}^0}{\partial T}
-\tau_{\lambda''}^0 v_{\lambda''}^c
\frac{\partial n_{\lambda''}^0}{\partial T}
\Big]
\\&\times\Big[\delta(\omega+\omega_{\lambda'}-\omega_{\lambda''})-\delta(\omega-\omega_{\lambda'}+\omega_{\lambda''})\Big]
\Big\},
\end{aligned}$}
\label{eq:gamma}
\end{equation}
where $\Phi_{-\lambda\lambda'\lambda''}$ is the cubic interaction matrix
element and $\omega\equiv\omega_\lambda$ is the probe frequency.
This response follows from the imaginary part of the one-loop retarded
self-energy evaluated with the first-order scattering-partner occupation corrections.
The nonequilibrium occupations change the damping without introducing a
new interaction vertex.
The correction and equilibrium lifetime are evaluated at the same single-mode RTA level.
Equation~\eqref{eq:gamma} involves only cubic force constants and
equilibrium phonon properties, both available from standard
lattice-dynamics calculations
\cite{WOS:000352075300002, WOS:000358925200003, WOS:000337929400021, WOS:000384537600001}.

Each channel in Eq.~\eqref{eq:gamma} is weighted by scattering-partner
velocities, equilibrium lifetimes, and temperature derivatives of the equilibrium occupations.
These signed weights determine the probe linewidth correction, so its sign
is not set by the probe velocity alone.
The equilibrium linewidth does not retain these weights, so
$\gamma_\lambda^c$ contains additional information about the anharmonic channels.

The corresponding relaxation time obeys
$\tau_\lambda=[2\Gamma_\lambda(\omega_\lambda)]^{-1}$ within the RTA.
With $\Gamma_\lambda^0=[2\tau_\lambda^0]^{-1}$, the corrected linewidth
and relaxation time are
\begin{equation}
\Gamma_\lambda(\nabla T)=\Gamma_\lambda^0-\gamma_\lambda^c\nabla_c T,
\label{eq:Gamma}
\end{equation}
\begin{equation}
\tau_\lambda(\nabla T)=\tau_\lambda^0+
\frac{\gamma_\lambda^c}{2[\Gamma_\lambda^0]^2}\nabla_c T,
\label{eq:tau}
\end{equation}
to first order in the gradient, with a sum over the repeated Cartesian
index $c$.
We denote the single-mode corrections by
$\delta\Gamma_\lambda=\Gamma_\lambda-\Gamma_\lambda^0$ and
$\delta\tau_\lambda=\tau_\lambda-\tau_\lambda^0$.
At this order, the expansion parameter satisfies
$|\gamma_\lambda^c\nabla_c T|/\Gamma_\lambda^0
=|\delta\tau_\lambda|/\tau_\lambda^0$.
The last column of Table~\ref{tab:tau} lists the corresponding signed
relative relaxation-time corrections.
Equation~\eqref{eq:Gamma} gives the leading heat-current correction to
the intrinsic linewidth, rather than to a transport coefficient.
The gradient-linear relaxation-time correction in Eq.~\eqref{eq:tau} is
absent when the lifetime is fixed at its equilibrium value.
Its feedback contributes to $n_\lambda^2$, while $n_\lambda^1$ is evaluated
with the equilibrium lifetime.

We use $J$ for the signed heat current along a fixed reference direction.
The relation
$\gamma_\lambda^c(-\mathbf q)=-\gamma_\lambda^c(\mathbf q)$
(see Symmetry considerations) then gives the paired linewidth contrast
\begin{equation}
\Delta\Gamma_\lambda(J)
=\Gamma_\lambda(J)-\Gamma_{-\lambda}(J)
=-2\gamma_\lambda^c(\mathbf q)\,\nabla_c T,
\label{eq:splitting}
\end{equation}
where the first equality defines the linewidth difference between time-reversed modes.
Reversing the heat current reverses the temperature gradient, giving
\begin{equation}
\Delta\Gamma_\lambda(-J)=-\Delta\Gamma_\lambda(J).
\label{eq:reversal}
\end{equation}
For a nonzero response, one member of a time-reversed pair broadens while the other narrows.
The first-order linewidth correction vanishes at $\mathbf q=0$, where oddness requires
$\gamma_\lambda^c=0$.

\rnhead{Symmetry considerations}\label{sec:symmetry}
Microscopic time-reversal symmetry gives equal equilibrium linewidths
at opposite wave vectors \cite{WOS:000201554900017}.
A steady heat current drives the phonon distribution away from equilibrium.
The first-order occupation correction for each scattering partner is odd
under reversal of its wave vector.
The coefficient $\gamma_\lambda^c$ is therefore odd in the probe wave vector
(see Sec.~II of the SM~\cite{SupplementalMaterial}).
Within this first-order expansion, Eq.~\eqref{eq:Gamma} gives
$\Gamma_\lambda(\mathbf q;\nabla T)=\Gamma_\lambda(-\mathbf q;-\nabla T)$.
This relation compares steady states with opposite heat currents at
the same local temperature.
At fixed drive, $\Gamma_\lambda(\mathbf q;\nabla T)\neq
\Gamma_\lambda(-\mathbf q;\nabla T)$ whenever
$\gamma_\lambda^c\nabla_c T\neq 0$.
The nonreciprocity therefore belongs to the heat-carrying state, not to
magnetic order or spontaneous time-reversal-symmetry breaking in the equilibrium crystal.
Symmetry fixes the oddness of $\gamma_\lambda^c$, while the scattering
channels determine its sign and magnitude for each mode.

This mode-resolved response differs from the intrinsic linear phonon
thermal Hall effect.
The latter involves the phonon Berry curvature, which is odd in
$\mathbf q$ under time reversal.
Its contributions cancel in the Brillouin-zone sum for a
time-reversal-invariant crystal
\cite{WOS:000232443400041, WOS:000236969700038, WOS:000308689600002, WOS:000286731000012, WOS:000503245200012}.
The linewidth contrast in Eq.~\eqref{eq:splitting} instead compares
paired modes without performing this sum.
It requires neither Berry curvature nor phonon angular momentum
\cite{WOS:000448463300012, WOS:000331958800003}, and no magnetic field is needed.
The response is therefore allowed in centrosymmetric nonmagnetic
crystals within a first-order expansion about local equilibrium.

\rnhead{First-principles results}\label{sec:results}
We evaluate Eq.~\eqref{eq:gamma} for MgO, diamond, and TaAs.
MgO and diamond are centrosymmetric ionic and covalent crystals,
respectively. They test the mechanism without magnetic order or broken
inversion symmetry in the equilibrium structure.
TaAs is a noncentrosymmetric Weyl semimetal (see Fig.~\href{supplemental_material.pdf\#figure.2}{S2} in the
SM~\cite{SupplementalMaterial})
\cite{WOS:000351507700001,Huang2015TaAs,WOS:000359092900034,WOS:000360709200013}.
All three materials are nonmagnetic and time-reversal invariant at equilibrium.
For TaAs, we isolate the phonon--phonon channel rather than predict the
total linewidth, which can also depend on electrons.

We obtain harmonic and cubic force constants using finite-displacement
calculations with phonopy and phono3py
\cite{Togo2023Implementation,WOS:000352075300002}.
Computational details and comparisons with equilibrium phono3py results are given in
Sec.~V of the SM~\cite{SupplementalMaterial}.
The quoted equilibrium lifetimes describe the intrinsic three-phonon channel.
We use numerical Gaussian smearing widths of $\sigma=0.1$ THz
for MgO and TaAs and $1.0$ THz for diamond.

To compare paired modes, we define the dimensionless linewidth asymmetry
\begin{equation}
\mathcal A_\lambda
=\frac{\Gamma_\lambda-\Gamma_{-\lambda}}
{\Gamma_\lambda+\Gamma_{-\lambda}}
=-\frac{\delta\tau_\lambda}{\tau_\lambda^0},
\label{eq:A}
\end{equation}
to first order in $\nabla T$.
Table~\ref{tab:tau} lists the equilibrium lifetimes, $\gamma_\lambda^x$,
and relaxation-time corrections for six selected modes at $T=50$ K and
$\nabla_x T=10^3$ K/m.
Each row gives one member of a $\pm\mathbf q$ pair.
The correction for the other member follows from the oddness of
$\gamma_\lambda^c$.

\begin{figure}[!b]
\centering
\includegraphics[width=0.85\linewidth]{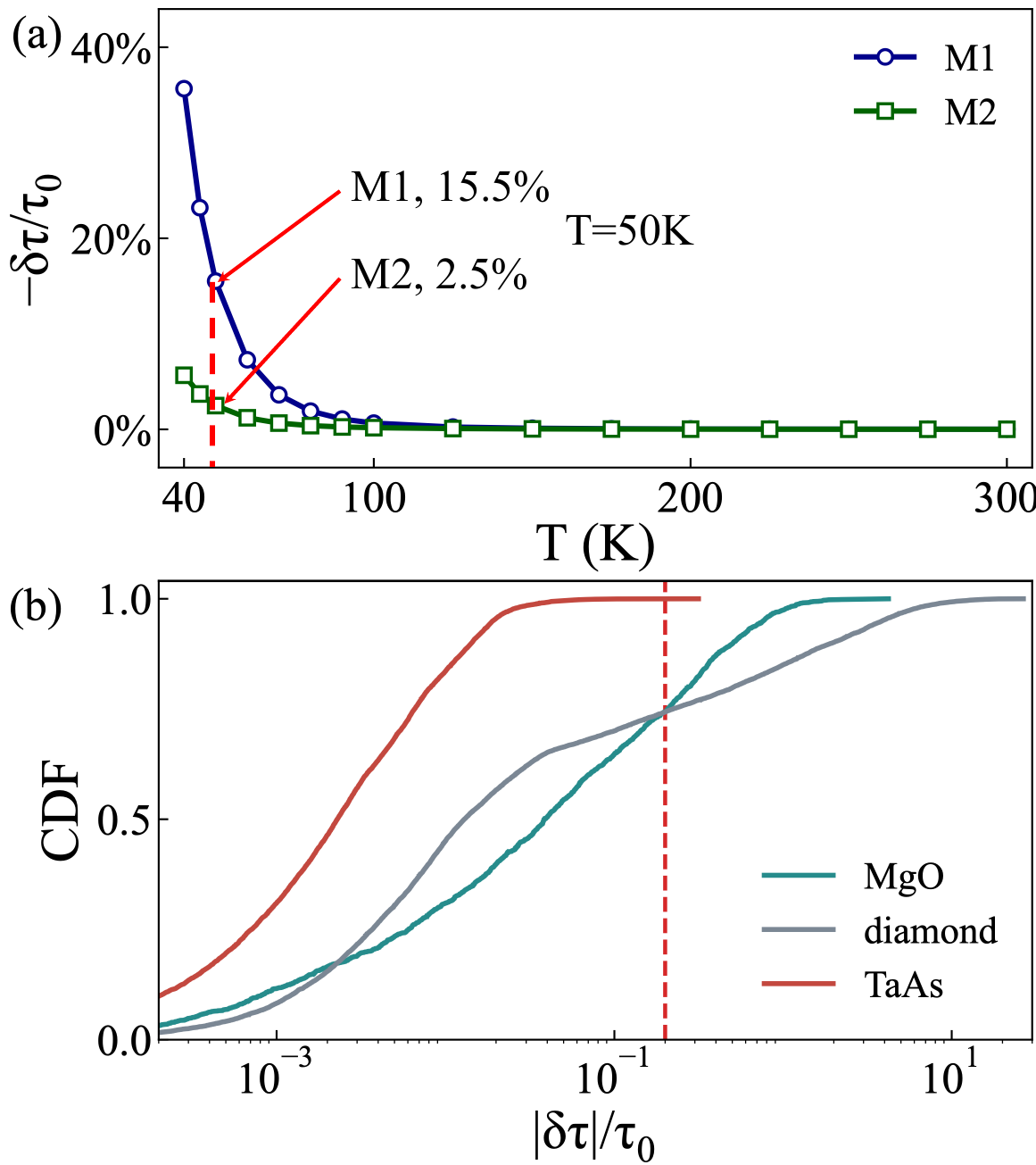}
\caption{Temperature dependence and mode distribution of the response at
$|\nabla T|=10^3$ K/m. (a) Temperature dependence of MgO M1 and M2.
(b) Cumulative distributions of $|\delta\tau_\lambda|/\tau_\lambda^0$
at $T=50$ K for MgO, TaAs, and diamond. The red dashed line marks the
$0.2$ first-order reporting threshold. TaAs is shown for the
phonon--phonon channel.}
\label{fig:results_summary}
\end{figure}

Figure~\ref{fig:results_summary}(a) compares MgO M1 and M2 over
$40$--$300$ K, with the $50$ K values listed in Table~\ref{tab:tau}.
For both modes, the magnitudes of the relative corrections and the
equilibrium lifetimes decrease as temperature increases. We assess the relative size of the first-order corrections using the
cumulative distributions of $|\delta\tau_\lambda|/\tau_\lambda^0$
over the full meshes [Fig.~\ref{fig:results_summary}(b)].
As a practical criterion, we regard the expansion as controlled when
$|\delta\tau_\lambda|/\tau_\lambda^0\lesssim 0.2$, as indicated by the red dashed line.
At $10^3$ K/m, this condition holds for $99.99\%$ of the TaAs entries
and about $74\%$ of the entries in both diamond and MgO.
All modes in Table~\ref{tab:tau} satisfy this criterion.
The smearing-width dependence of these corrections is reported in
Sec.~VI of the SM~\cite{SupplementalMaterial}.

The quantitative predictions remain restricted to these finite-$\mathbf q$
modes at the stated numerical smearing widths.
The magnitude of the relative correction increases toward the smallest controlled wave vectors,
whereas symmetry requires the linewidth correction to vanish at $\mathbf q=0$.
We therefore do not extrapolate below the controlled range.
For MgO, the closest controlled wave vector is $|\mathbf q|\approx0.19$
in reduced reciprocal-lattice coordinates.
At the next smaller mesh point, $|\mathbf q|\approx0.125$ in the same units,
the correction becomes sensitive to the smearing width.
This limits the connection to smaller-wave-vector Brillouin and
ultrasonic measurements, as discussed below.

The corrections are mode selective and can have either sign, as expected
from the signed partner contributions in Eq.~\eqref{eq:gamma}.
Table~\ref{tab:tau} also distinguishes the response coefficient from the
relative linewidth contrast. MgO M2 has a larger $|\gamma^x|$ than M1,
but a smaller relative-correction magnitude because its equilibrium linewidth is
larger. A large response coefficient therefore need not produce the
largest relative linewidth change. For opposite
wave vectors, Eq.~\eqref{eq:splitting} gives the linewidth ratio
\begin{equation}
\frac{\Gamma_\lambda(+\mathbf q)}{\Gamma_\lambda(-\mathbf q)}
=\frac{1+\mathcal A_\lambda}{1-\mathcal A_\lambda}
\simeq 1-\frac{2\gamma_\lambda^x\nabla_x T}{\Gamma_\lambda^0},
\label{eq:ratio}
\end{equation}
where the final expression retains terms linear in the small relative linewidth correction
$|\delta\Gamma_\lambda|/\Gamma_\lambda^0\ll1$.

\begin{figure}[!t]
\centering
\includegraphics[width=\linewidth]{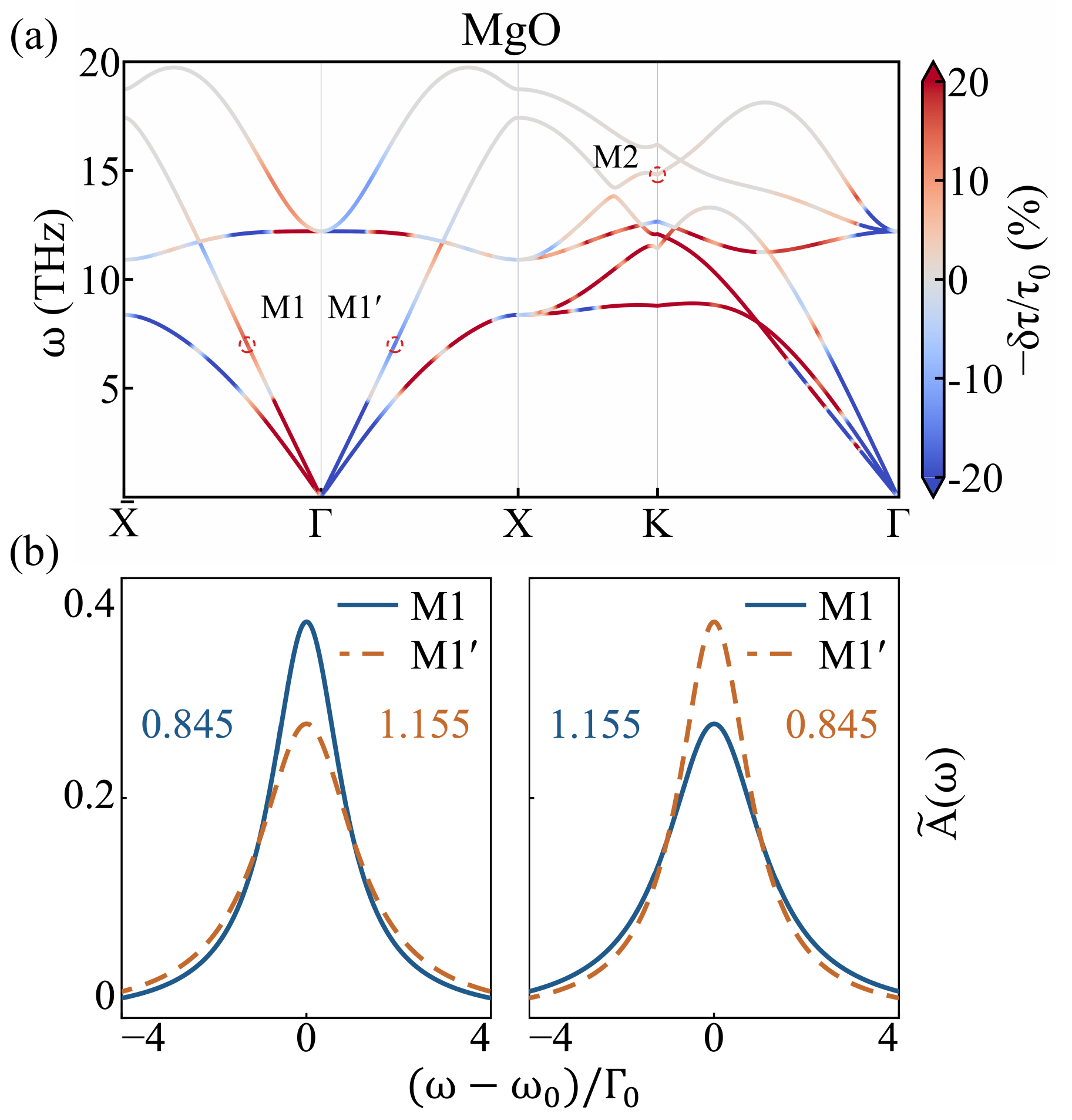}
\caption{Wave-vector-resolved response and heat-current reversal in MgO.
(a) First-principles dispersion along
$\bar{X}$--$\Gamma$--$X$--$K$--$\Gamma$, colored by
$\mathcal A_\lambda=-\delta\tau_\lambda/\tau_\lambda^0$ (\%).
The color scale saturates at $\pm20\%$, and values outside this range
use the corresponding endpoint colors.
M1 and M1$^\prime$ mark a time-reversed pair, and M2 marks a separate
controlled mode.
(b) Area-normalized line shapes reconstructed from the equilibrium phono3py
self-energy and the calculated on-shell linewidth corrections.
The two panels compare opposite heat-current directions, which exchange
the broad and narrow profiles. Peak centers are aligned to isolate the
intrinsic linewidth contrast. The profiles do not include the instrumental response.}
\label{fig:signatures}
\end{figure}

Figure~\ref{fig:signatures}(a) shows the wave-vector-resolved linewidth
asymmetry in MgO.
The paired M1 and M1$^\prime$ modes have opposite-sign linewidth corrections,
as expected from the wave-vector-odd response.
Their line shapes in Fig.~\ref{fig:signatures}(b) exchange broad and narrow
profiles upon heat-current reversal.
For the selected MgO M1 mode, the dimensionless linewidth asymmetry is
$\mathcal A_\lambda=15.5\%$.
The unexpanded ratio $(1+\mathcal A_\lambda)/(1-\mathcal A_\lambda)$ in
Eq.~\eqref{eq:ratio}, formed from the first-order linewidths, is approximately $1.37$.

\rnhead{Spectroscopic implications}
The predicted asymmetry concerns the intrinsic peak width, not the
occupation-dependent peak area or apparent broadening from spatial
temperature variations.
The relevant comparison is between the linewidths of time-reversed modes
at fixed local temperature, repeated after current reversal.
Extracting the intrinsic linewidth requires separating it from peak-area
and peak-position changes, Bose occupation factors, and instrumental broadening.
Assessing experimental resolution requires the absolute linewidth
difference and instrumental response at the calculated wave vectors, not
the relative contrast alone.
Figure~\ref{fig:signatures}(b) illustrates the intrinsic width exchange,
not a simulated measurement.

The linewidth asymmetry also produces directional attenuation of coherent
phonon wave packets.
For a mode at wave vector $\mathbf q$, the spatial amplitude-attenuation
coefficient is $\alpha_\lambda=\Gamma_\lambda/|v_\lambda|$,
where $|v_\lambda|$ is the group speed.
Equation~\eqref{eq:splitting} therefore gives
\begin{equation}
\alpha_\lambda(+\mathbf q)-\alpha_\lambda(-\mathbf q)
=-2\gamma_\lambda^c\nabla_c T/|v_\lambda|,
\label{eq:attenuation}
\end{equation}
to first order in the gradient.
The relation applies to the modes for which the linewidth calculation is
controlled. Brillouin and ultrasonic measurements probe smaller wave
vectors than those quantified here, so their signal magnitude remains
outside the present predictions.

\rnhead{Summary}\label{sec:conclusions}
In summary, a steady phonon heat current can make intrinsic anharmonic
damping nonreciprocal even in centrosymmetric nonmagnetic crystals.
The response arises from unequal partner occupations, not unequal
squared interaction matrix elements.
Its mode-resolved coefficient retains signed channel weights absent
from the equilibrium linewidth.
The MgO and diamond examples establish this response within the RTA
at the stated numerical smearing widths.
Current reversal exchanges the paired linewidth corrections, providing
a spectral criterion distinct from population-induced intensity changes.
The electron--phonon example in Sec.~IV of the SM~\cite{SupplementalMaterial}
illustrates analogous occupation feedback in electron damping under an electric field.

The predicted linewidth contrast offers a spectroscopic signature of
occupation feedback in phonon damping.
Momentum-resolved spectroscopy could test this prediction by comparing
the linewidths of time-reversed modes under heat-current reversal at fixed local temperature.
The gradient-linear phonon damping correction may contribute to nonlinear
phonon Hall transport in suitable symmetry settings
\cite{Im2022NonlinearPhononHall,Luo2023NonlinearPhononHall}.
Incorporating this feedback into bosonic transport theory
\cite{Kuwabara2026BosonicTransport} would allow its contribution to
the symmetry-allowed transverse heat current to be assessed.

\rnhead{Acknowledgments}
This work was supported by the National Key Research and Development Program
of the Ministry of Science and Technology of China (Grant No. 2025YFA1411303)
and the National Natural Science Foundation of China (Grant Nos. 12595333 and 12604064).

\rnhead{Data availability}
The numerical inputs required to reproduce the results are provided in the
SM~\cite{SupplementalMaterial}. Full-grid data and analysis scripts are available from the
corresponding authors upon reasonable request.

\nocite{WOS:A1993MF90100084,WOS:A1996VT67500040,WOS:A1994QB02200016,WOS:A1996VP22500044}

%

\end{document}